\documentclass[aps,prc,twocolumn,showpacs,groupedaddress,floatfix]{revtex4}

\usepackage{graphicx}
\usepackage{dcolumn}
\usepackage{amsfonts}

\begin{document}

\title{$^{64}$Ni+$^{64}$Ni fusion reaction calculated with the density-constrained
time-dependent Hartree-Fock formalism}

\author{A.S. Umar and V.E. Oberacker}
\affiliation{Department of Physics and Astronomy, Vanderbilt University,
             Nashville, Tennessee 37235, USA}

\date{\today}


\begin{abstract}
We study fusion reactions of the $^{64}$Ni+$^{64}$Ni system using the
density-constrained time-dependent Hartree-Fock (TDHF) formalism. In this formalism the
fusion barriers are directly obtained from TDHF dynamics. In addition, we incorporate the
entrance channel alignments of the slightly deformed (oblate) $^{64}$Ni nuclei due to dynamical
Coulomb excitation. We show that alignment leads to a fusion barrier distribution
and alters the naive picture for defining which energies are actually sub-barrier.
We also show that core polarization effects could play a significant role in
fusion cross section calculations.

\end{abstract}
\maketitle

\section{\label{sec:intro}Introduction}
Radioactive ion-beam facilities enable us to study the structure and reactions of
exotic nuclei, in particular the physics properties of the ``terra incognita''
of neutron-rich isotopes~\cite{DOE02}. One important aspect of these studies is a
detailed investigation of the heavy-ion fusion process of exotic nuclei. This is crucial
not only for superheavy element formation, but it will also lead to
a better understanding of the effective $N-N$ interactions in neutron-rich nuclei
and of enhanced correlations present in these many-body systems.

Recently, fusion evaporation cross sections for the $^{64}$Ni+$^{64}$Ni system have been
measured down to the $10$~nb level~\cite{Ji04}. This experiment confirmed and improved the earlier data~\cite{BS82}
for the same system and it extended the data to extreme sub-barrier energies,
thus providing a challenge for the
theoretical understanding of the fusion process between two open-shell nuclei.
The primary observation was a \emph{hindrance} of fusion in the $^{64}$Ni+$^{64}$Ni system
at extreme sub-barrier energies in comparison to reactions involving other Nickel
isotopes such as the $^{58}$Ni+$^{58}$Ni system.
Earlier coupled-channels calculations~\cite{Esb04,Esb05}
failed to reproduce the data at the extreme sub-barrier energies.

Various hypotheses were developed for the explanation of the fusion hindrance phenomenon.
In Ref.~\cite{Esb06} the hindrance was attributed to the differing stiffness of Nickel isotopes
due to nuclear structure effects. An
excellent coupled-channels fit to the data was obtained by supplementing the effective
$N-N$ force used in the double-folding potential with a repulsive core to account
for the nuclear incompressibility effects at the nuclear overlap, thus leading to a
shallow potential pocket. On the other hand, Refs.~\cite{IH07a,IH07b} suggest that at such
low energies the inner turning point of the heavy-ion potential is smaller
than the touching point $r_t=R_1+R_2$. Thus the
validity of the frozen-density approximation used in Ref.~\cite{Esb06} becomes questionable.
These authors have proposed a two-step model for fusion in which the effects of
neck formation are approximately included~\cite{IH07a}.

The theoretical analysis of the fusion data generally involves the determination
of a phenomenological ion-ion potential such as the Bass model~\cite{Ba74,Ba80},
the proximity potential~\cite{BR77,RV78,SG84,BH78}, or potentials obtained via
the double-folding method~\cite{SL79,BB77,RO83a,RO83b}. Subsequently, the actual
fusion cross section is calculated by either using barrier penetration
models~\cite{Ba80,TB84,RO83a,BT98}, or the coupled-channel method~\cite{LP84,RP84,HR99,Esb04,Esb05}.
The latter includes various excitations of the
target and/or projectile using the coupled-channel formalism~\cite{HR99,Esb04}, as well as
the inclusion of neutron transfer, and can be consistently applied at energies above
and below the barrier~\cite{BT98}. Effectively, the inclusion of each additional excitation
leads to a modification of the original inert core ion-ion potential, resulting in a series
of effective barriers.
One common physical assumption used in many of these
calculations is the use of the frozen density or the sudden
approximation. In this approximation the nuclear densities
are unchanged during the computation of the ion-ion potential as a function
of the internuclear distance. Furthermore, the effects included in channel couplings
are usually based on the static properties of the participating nuclei, which may
accurately represent the early stages of the collision process, but are expected
to change as the two ions strongly interact. While these methods provide a useful
and productive means for quantifying multitudinous reaction data it is desirable to
include dynamical effects and make contact with the microscopic theories of nuclear
structure and reactions.

Recently, we have developed a microscopic approach for calculating heavy-ion interaction
potentials which incorporates all of the dynamical entrance channel effects included in the
time-dependent Hartree-Fock (TDHF) description of the collision process~\cite{UO06b}.
These effects include the neck formation, particle exchange, internal excitations,
and deformation effects to all order, as well as the effect of nuclear alignment
for deformed systems. The method is based on the TDHF
evolution of the nuclear system coupled with density-constrained Hartree-Fock
calculations to obtain the ion-ion interaction potential. Preliminary calculations
for the $^{64}$Ni+$^{132}$Sn system highlighted the importance of dynamical
deformation effects~\cite{UO06d,UO07a}. Here we give a completed study of fusion cross
sections using this formalism.

In the next section we will summarize some theoretical aspects of the density-constrained
TDHF theory along with methods to incorporate dynamical alignment into our calculations,
as well as the method used to calculate cross sections from the resulting barriers.
In Section~\ref{sec:results}
we present interesting aspects of the reaction dynamics and compare our results with
experiment and other calculations. In section~\ref{sec:summary} we summarize our
conclusions.

\section{\label{sec:formal}Theoretical Methods}

\subsection{\label{sec:dctdhf} Density-constrained TDHF method}
For the calculations of dynamical potential barriers for the $^{64}$Ni+$^{64}$Ni
system we have used the density-constrained TDHF (DC-TDHF) method.
Further details of the method can be found in Ref.~\cite{UO06b}, here we give
a short overview.

The {\it density constraint} is a novel numerical method that was developed in
the mid 1980's~\cite{CR85,US85} and was used to provide a microscopic
description of the formation of shape resonances in light systems~\cite{US85}.
In this approach the TDHF time-evolution takes place with no restrictions.
At certain times during the evolution the instantaneous density is used to
perform a static Hartree-Fock minimization while holding the total density constrained
to be the instantaneous TDHF density. In essence, this provides us with the
TDHF dynamical path in relation to the multi-dimensional static energy surface
of the combined nuclear system. Since we are constraining the total density
all moments are simultaneously constrained.
The numerical procedure for implementing this constraint and the method for
steering the solution to $\rho_{\mathrm{TDHF}}(\mathbf{r},t)$ is discussed in Refs.~\cite{CR85,US85}.
The convergence property is as good if not better than in the traditional
constrained Hartree-Fock
calculations with a constraint on a single collective degree of freedom.

In Ref.~\cite{UO06b} we have shown that the ion-ion interaction potential is
given by
\begin{equation}
V(R)=E_{\mathrm{DC}}(R)-E_{\mathrm{A_{1}}}-E_{\mathrm{A_{2}}}\;,
\label{eq:vr}
\end{equation}
where $E_{\mathrm{DC}}$ is the density-constrained energy at the instantaneous
separation $R(t)$, while $E_{\mathrm{A_{1}}}$ and $E_{\mathrm{A_{2}}}$ are the binding energies of
the two nuclei obtained with the same effective interaction. We would like to
emphasize again that this procedure does not affect the TDHF time-evolution and
contains no {\it free parameters} or {\it normalization}. In practice,
TDHF runs are initialized with energies above the Coulomb barrier and in
Ref.~\cite{UO06b} we have shown that there is no appreciable energy dependence
to the barriers obtained via the density-constrained TDHF method.
The separation coordinate $R$ is the distance
between the centers of mass of the two nuclei.

\subsection{\label{sec:align} Fusion for deformed nuclei}
\begin{figure}[!htb]
\includegraphics*[scale=0.50]{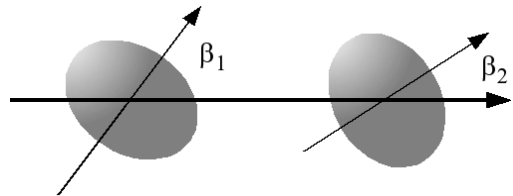}
\caption{\label{fig:align_schem} Shown are the orientation angles $\beta_1$ and $\beta_2$ of the oblate Ni
nuclei with respect to the collision axis.}
\end{figure}

The heavy-ion interaction potential between two deformed nuclei depends on the distance
vector between their centers-of-mass, ${\bf R}$, and on the relative orientation of their
intrinsic principal axis systems which may be described in terms of three Euler angles
$(\alpha,\beta,\gamma)$ per nucleus, i.e. in the most general case we have
\begin{equation}
V = V ( {\bf R},\alpha_1,\beta_1,\gamma_1,\alpha_2,\beta_2,\gamma_2 )\;.
\end{equation}
The expression for $V$ can be simplified
if the intrinsic nuclear density distributions are axially symmetric;
in this case, the potential does not depend on the Euler angles $\gamma_1,\gamma_2$
which describe rotations about the symmetry axes. If we put, for convenience,
the distance vector in $z$-direction, ${\bf R}= R {\bf e}_z$, the potential
between two deformed axially symmetric nuclei has the structure
\begin{equation}
V = V ( R,\beta_1,\beta_2, \Delta \alpha)\;.
\end{equation}
The heavy-ion interaction potential is calculated with the density-constrained TDHF method
for a given set of orientation angles $\beta_1,\beta_2, \Delta \alpha$. Fortunately,
test calculations using the double-folding method described in Ref.~\cite{RO83b}
reveal that the dependence on the Euler angle $\Delta \alpha$ is negligible in our case
so we put $\Delta \alpha = 0$. In Fig.~\ref{fig:align_schem} we show the definition of
the angles $\beta_{1}$ and $\beta_{2}$ for two oblate nuclei.

In order to calculate the total fusion cross section at energy $E_{\mathrm{c.m.}}$,
we first consider the partial cross section for given initial orientations
$(\beta_1,\beta_2)$ of the two nuclei which is given by
\begin{equation}
\label{eq:fusion_beta}
\sigma(E_{\mathrm{c.m.}},\beta_1,\beta_2)=\frac{\pi}{k_0^2}\sum_{L=0}^{\infty}
          (2L+1) T_L(E_{\mathrm{c.m.}},\beta_1,\beta_2)\;,
\end{equation}
with $k_0=\sqrt{2\mu E_{\mathrm{c.m.}}}$.
The fusion barrier penetrabilities $T_L(E_{\mathrm{c.m.}},\beta_1,\beta_2)$
are obtained by numerical integration of the two-body Schr\"odinger equation
using the {\it incoming wave boundary condition} (IWBC) method~\cite{Raw64,LP84}
\[
\left[ \frac{-\hbar^2}{2\mu}\frac{d^2}{dR^2}+\frac{L(L+1)\hbar^2}{2\mu R^2}+V(R,\beta_1,\beta_2)-E\right]\psi=0\;,
\]
where the quantity $V(R,\beta_1,\beta_2)$ denotes the heavy-ion potential obtained via
the density-constrained TDHF method. For the numerical implementation we have followed the procedure
for the coupled-channel code CCFUL described in Ref.~\cite{HR99}.
IWBC assumes that once the minimum of the potential is reached fusion will
occur. In practice, the Schr\"odinger equation is integrated from the potential
minimum, $R_\mathrm{min}$, where only an incoming wave is assumed, to a large asymptotic distance,
where it is matched to incoming and outgoing Coulomb wavefunctions. The barrier
penetration factor, $T_L(E_{\mathrm{c.m.}},\beta_1,\beta_2)$ is the ratio of the
incoming flux at $R_\mathrm{min}$ to the incoming Coulomb flux at large distance.
\begin{figure}[!htb]
\includegraphics*[scale=0.40]{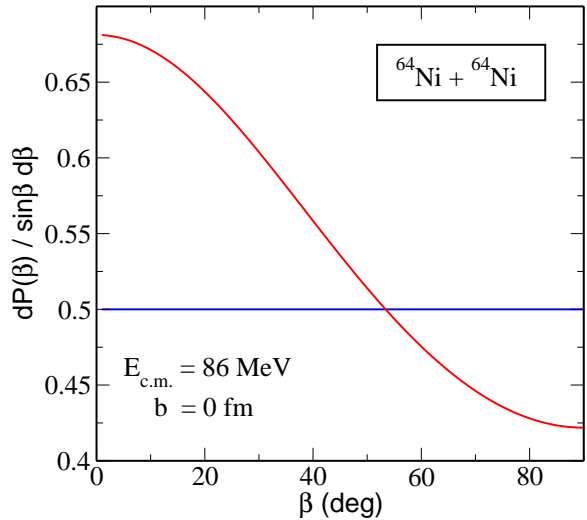}
\caption{\label{fig:align_ni_sn} (Color online) Dynamic alignment due to Coulomb
excitation of $^{64}$Ni. Shown is the orientation probability
as a function of the Euler angle $\beta$ in a central collision at internuclear
distances $R=1500$~fm (blue curve) and at $R=16$~fm (red curve).}
\end{figure}

Once the partial fusion cross sections (\ref{eq:fusion_beta}) for given orientation angles
$(\beta_1,\beta_2)$ have been calculated,
we have to take an average over all initial angular orientations of both nuclei
\begin{eqnarray*}
\label{eq:fusion}
\sigma(E_{\mathrm{c.m.}}) & = & \int_0^\pi \mathrm{sin}(\beta_1) d\beta_1\; \int_0^\pi \mathrm{sin}(\beta_2) d\beta_2
\nonumber \\
&\times& \frac{d^2P(E_{\mathrm{c.m.}},\beta_1,\beta_2)}{\mathrm{sin}(\beta_1) d\beta_1 \; \mathrm{sin}(\beta_2) d\beta_2}
\; \sigma(E_{\mathrm{c.m.}},\beta_1,\beta_2)\;,
\end{eqnarray*}
where $d^2P(E_{\mathrm{c.m.}},\beta_1,\beta_2)$ represents the alignment probability
for both deformed nuclei.
Details of the dynamic alignment formalism are presented in~\cite{UO06c}. We give here
a brief summary: For a given incident energy $E_{\mathrm{c.m.}}$ we carry out a semiclassical
Coulomb excitation calculation of the dominant collective levels of the deformed
nucleus. The energy levels and $EL$-transition matrix elements for $^{64}$Ni
are taken from experimental data~\cite{ENSDF}:  $E_{2+}=1.346$~MeV, $E_{4+}=2.610$~MeV
and $M(E2, 0+ \rightarrow 2+) = -27.0\ e~$fm$^2$  (oblate deformation).
The Coulomb excitation calculation starts at very large internuclear distances
(about $1500$~fm) when both nuclei may be presumed to be in their respective
ground states and stops at the ion-ion separation distance $R(t_0)$ (about $16$~fm). The Coulomb
excitation amplitudes determine the probability distribution of initial orientations.
Using the dominant monopole-multipole part of the Coulomb interaction, the
orientation probability factorizes as follows
\[
\frac{d^2P(E_{\mathrm{c.m.}},\beta_1,\beta_2)}{\mathrm{sin}(\beta_1) d\beta_1 \; \mathrm{sin}(\beta_2) d\beta_2}
= \frac{dP_1(E_{\mathrm{c.m.}},\beta_1)}{\mathrm{sin}(\beta_1) d\beta_1}
  \frac{dP_2(E_{\mathrm{c.m.}},\beta_2)}{\mathrm{sin}(\beta_2) d\beta_2} \;.
\]
In the special case of no preferential alignment, i.e. all initial orientation angles are
equally likely, this factor reduces to
\[
\left. \frac{d^2P(E_{\mathrm{c.m.}},\beta_1,\beta_2)}{\mathrm{sin}(\beta_1) d\beta_1 \;
 \mathrm{sin}(\beta_2) d\beta_2}\right|_{\mathrm{no align}}
\longrightarrow \frac{1}{4} \;\; .
\]
In Fig.~\ref{fig:align_ni_sn} we show the differential alignment probability
as a function of the Euler angle $\beta$ used in our calculations.

\section{\label{sec:results} Results}
We have carried out a number of TDHF calculations with accompanying
density constraint calculations to compute $V(R,\beta_1,\beta_2)$ given by Eq.~(\ref{eq:vr}).
A detailed description of our new three-dimensional unrestricted TDHF code
has recently been published in Ref~\cite{UO06}. The code was modified to self-consistently
generate initial states for $^{64}$Ni with different orientations.
For the effective interaction we have primarily used the Skyrme SLy5 force~\cite{CB98},
including all of the time-odd terms.
In this case the $^{64}$Ni nucleus is essentially oblate having a quadrupole moment of -0.45~b.
This is also confirmed by other calculations~\cite{DS04,LR99}.
\begin{figure}[!htb]
\includegraphics*[scale=0.36]{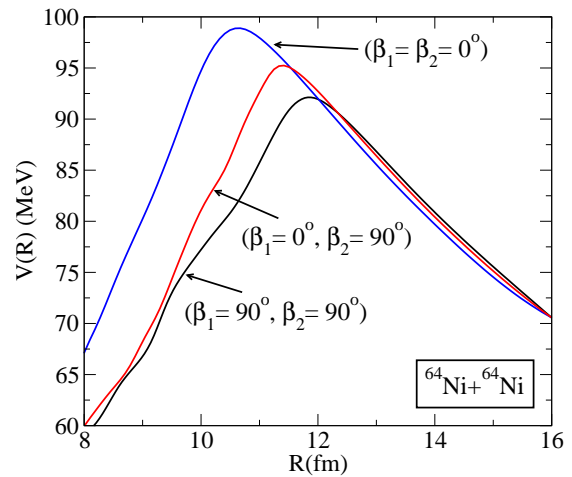}
\caption{\label{fig:vrb} (Color online) Potential barriers, $V(R,\beta_1,\beta_2)$, obtained from density-constrained
TDHF calculations for the $^{64}$Ni+$^{64}$Ni system. The Euler angles $\beta_1$ and $\beta_2$ indicate different
orientations of the deformed $^{64}$Ni nucleus.}
\end{figure}

All of our TDHF calculations were done at an initial energy of $E_{\mathrm{c.m.}}=98$~MeV
and separation $R(t_0)=16$~fm. As we have reported in Ref.~\cite{UO06b} the
potential barriers obtained from the density-constrained TDHF method are not
sensitive to the initial energy (above the barrier). We have
tested this again by running a few orientations at $112$~MeV and did not observe any
appreciable difference.
In Fig.~\ref{fig:vrb} we show the barriers obtained for limiting orientations of the
$^{64}$Ni nuclei.
We have also calculated these limiting barriers using other effective interactions, SkM$^*$~\cite{BQ82} and SLy4~\cite{CB98},
with essentially no difference. While most modern Skyrme parametrizations give essentially the same
potential barrier, using an older Skyrme force such as the SIII~\cite{BF75} interaction results in a higher barrier
and consequently lower fusion cross sections. This issue will be discussed later in the manuscript.

The physical picture which emerges from the barriers shown in Fig.~\ref{fig:vrb} is that the total
fusion cross section strongly depends on the deformation phase space. It also shows the fallacy of the
often-used statement that a certain energy is {\it sub-barrier}, which stems from spherical systems
that can be studied using a single barrier. For deformed systems this is dependent on the orientation
of the nuclei. For the $^{64}$Ni+$^{64}$Ni system the only truly sub-barrier energies are those below the
lowest potential barrier corresponding to the $\beta_1=\beta_2=90^{\circ}$ orientation, about $E_{\mathrm{c.m.}}=92$~MeV.
The fusion
cross sections corresponding to energies above the lowest barrier will be dominated by it since above-barrier
cross sections are much larger than the below-barrier ones.
\begin{figure}[!htb]
\includegraphics*[scale=0.40]{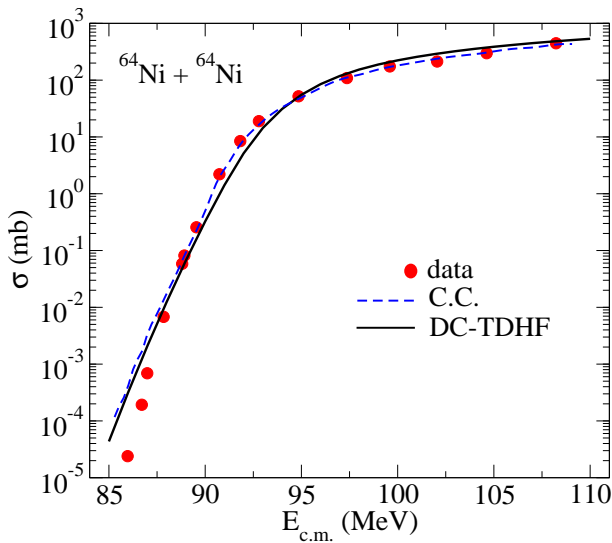}
\caption{\label{fig:xsection} (Color online) Total fusion cross section as a function of $E_{\mathrm{c.m.}}$.
Shown are the experimental data (filled red circles) and the coupled-channel calculation from Ref.~\cite{Ji04}
(blue dashed curve), and fusion cross sections calculated with the density-constrained TDHF method
using the SLy5 force (solid black line).}
\end{figure}

For the calculation of barrier distributions as a function of the orientations angles $\beta_1$ and $\beta_2$ we
have chosen an angular spacing of $\Delta\beta=10^{\circ}$. In principle this requires four hundred DC-TDHF calculations,
which would be very time consuming. However, one can show that some of the orientations are equivalent to each
other. One major assumption we have made is to assume the equality of the angular intervals $(0,\pi/2)$ and
$(\pi/2,\pi)$, which is not exactly correct when both nuclei are deformed. In order to assure that this
approximation does not effect the lowest energy cross sections (primarily determined by the lowest barrier)
we have explicitly calculated those angles that would have appreciable contribution at these energies. In
total we have computed twenty potential barriers corresponding to various orientations for the SLy5 force.
In principle even this may not be necessary since all of the barriers must fall between the limiting cases
shown in Fig.~\ref{fig:vrb}. Although actual calculations show that a constant angular interval $\Delta\beta$
does not always lead to equally spaced barriers such an extrapolation has a minimal effect on the actual
cross section calculations. We have confirmed this by generating such barriers from the limiting
barriers using a numerical averaging procedure and calculating the fusion cross section. This method was used for the calculation
of the cross sections for the SIII interaction.
\begin{figure}[!htb]
\includegraphics*[scale=0.40]{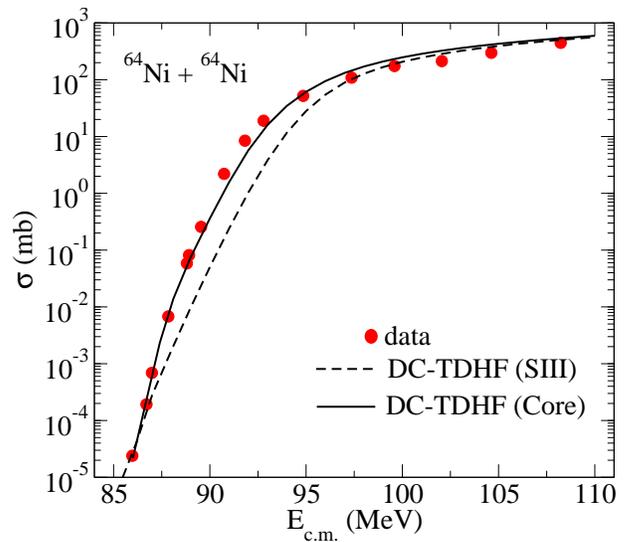}
\caption{\label{fig:SIII} (Color online) Total fusion cross section as a function of $E_{\mathrm{c.m.}}$.
Shown are the experimental data (filled red circles), density-constrained TDHF cross sections
using the SIII force (dashed black line), and the density-constrained TDHF cross sections using
the core orientation with the SLy5 force for the lowest energy cross sections.}
\end{figure}
\begin{figure*}[!hbt]
\begin{center}
\includegraphics*[scale=0.45]{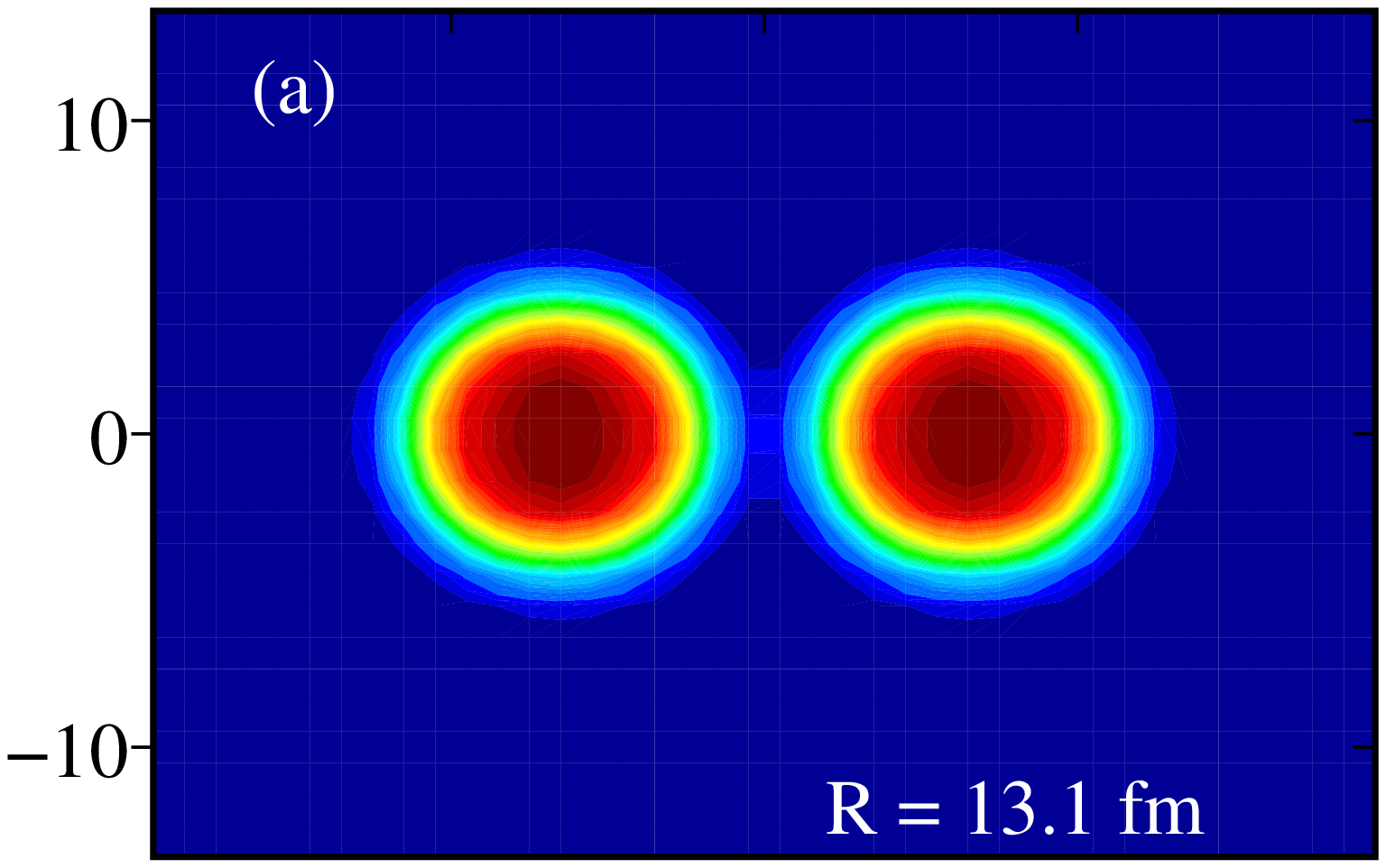}\hspace{-0.04in}
\includegraphics*[scale=0.45]{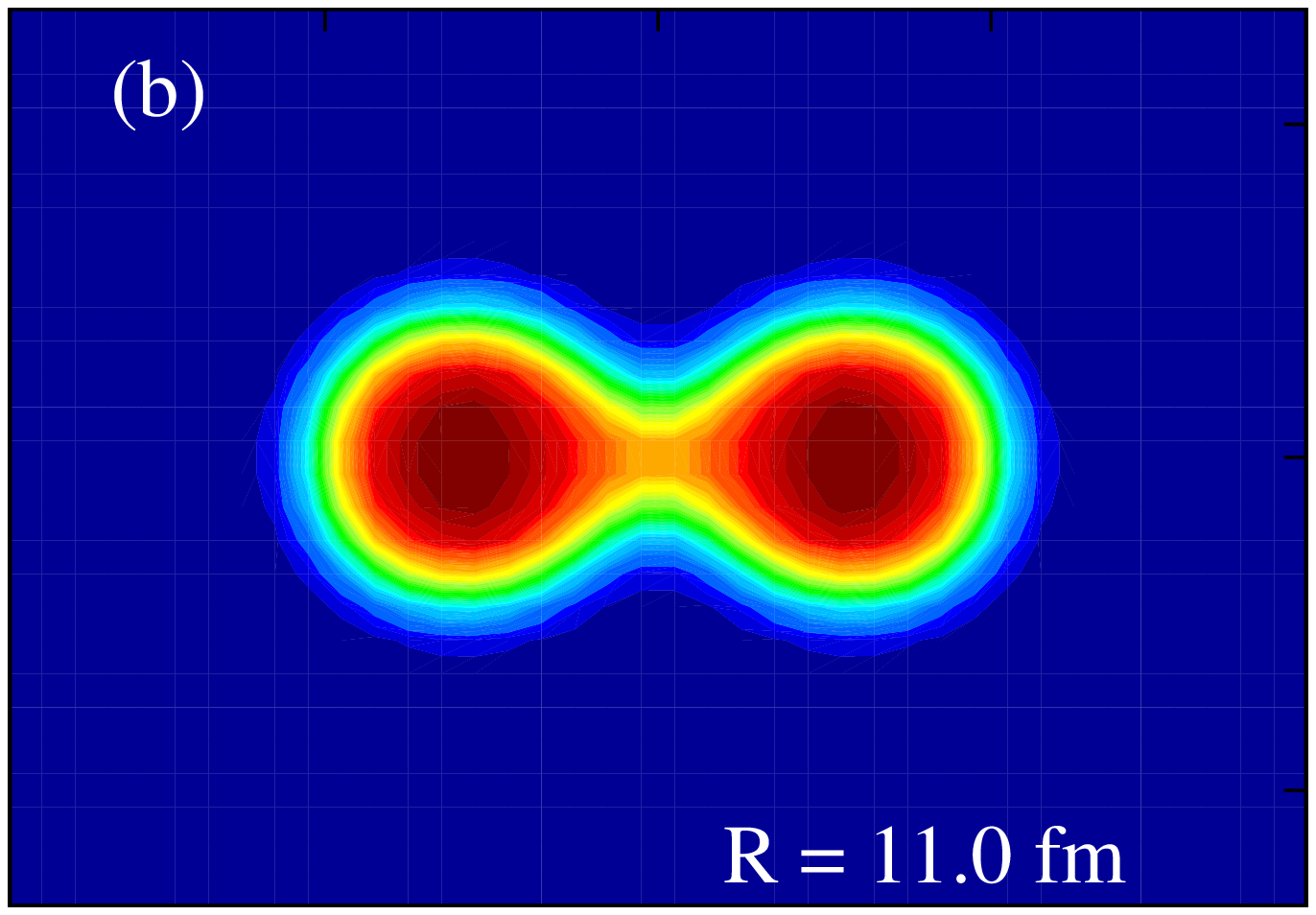}\\ \vspace{-0.02in} \hspace{0.43in}
\includegraphics*[scale=0.45]{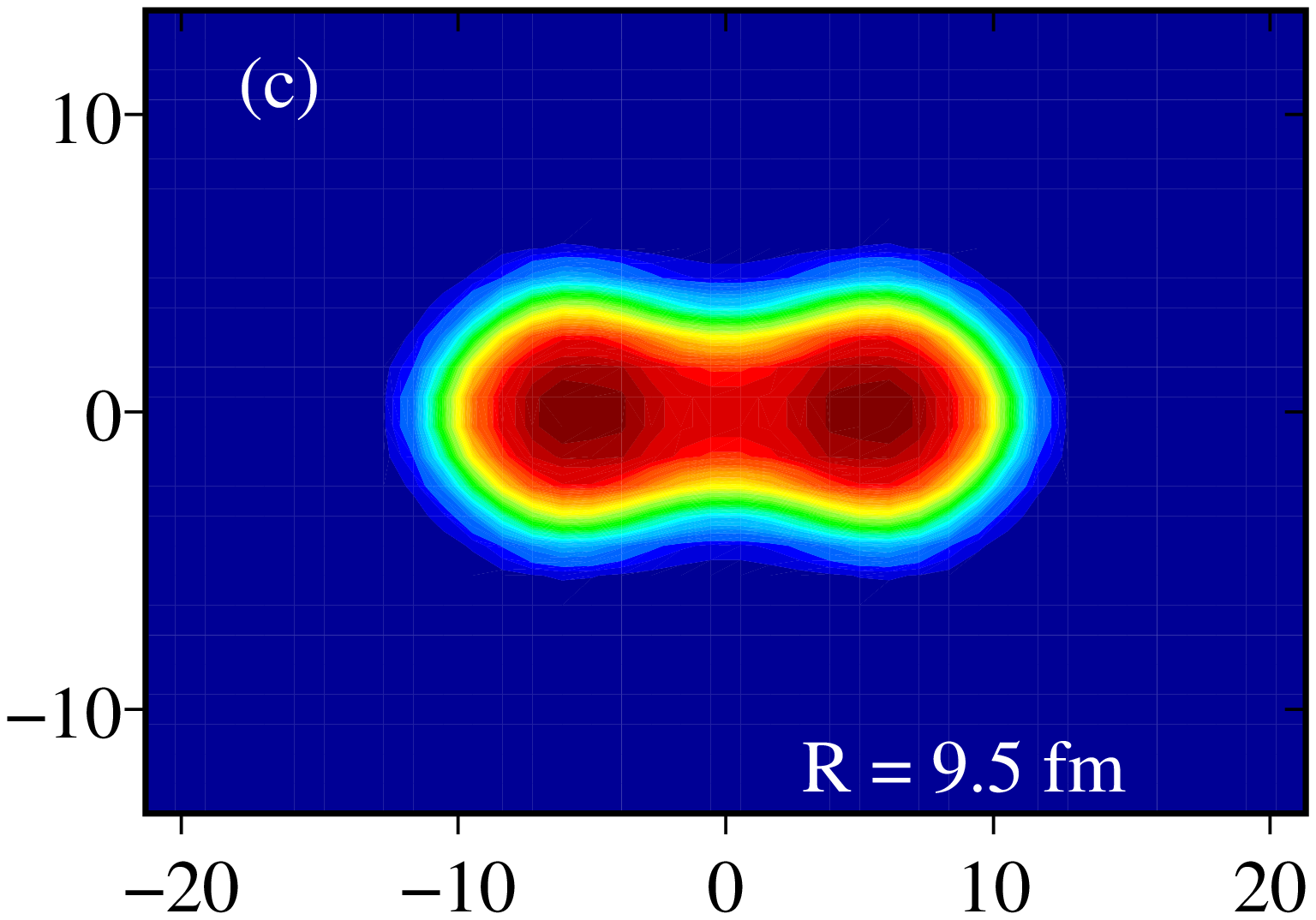}\hspace{-0.08in}
\includegraphics*[scale=0.45]{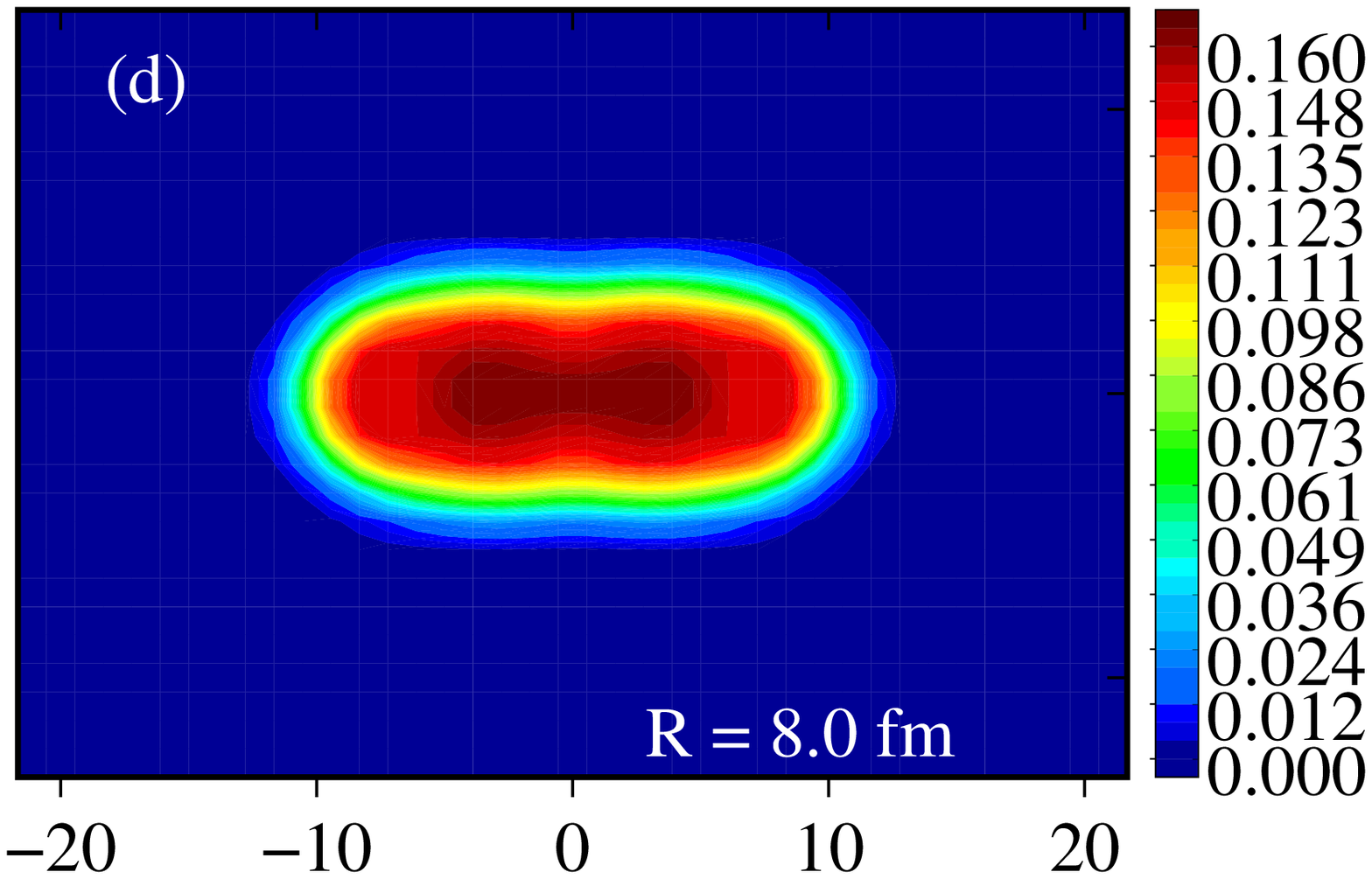}\hspace{-0.05in}
\caption{\label{fig:time-evol}(Color online) Shown are the density contours in the $x-z$ plane for the TDHF time-evolution
of the $^{64}$Ni+$^{64}$Ni system. Initially both nuclei are oriented with angles $\beta_1=\beta_2=90^{\circ}$
at a center-of-mass energy of $E_{\mathrm{c.m.}}=98$~MeV. The values for the ion-ion separation $R$
correspond to special points along the potential barrier at the lowest experimental energy of $E_{\mathrm{c.m.}}=86$~MeV;
(a) outer turning point, (b) inner turning point, (c) reorientation of the core, and (d) potential minimum.}
\end{center}
\end{figure*}

In Fig.~\ref{fig:xsection} we show the total DC-TDHF fusion cross section
as a function of the center-of-mass energy (solid black curve) using the SLy5 force. Also, shown are
the experimental data (filled circles), and the coupled-channels calculations of
Ref.~\cite{Ji04} (dashed blue line). Results for SLy4 and SkM$^*$ interactions
are indistinguishable from the SLy5 result. We observe that the DC-TDHF calculations, which contain
no parameters or normalization, accurately reproduce the fusion cross sections for all energies
except for energies $E_{\mathrm{c.m.}}\leq 87$~MeV. We believe that the small deviations at other
energies are largely due to the symmetry assumptions made in alignment averaging. As we stated earlier this
is not the case for the lowest energies as these were explicitly done without symmetry assumptions.

In Ref.~\cite{Esb06} the hindrance of the fusion cross section at deep sub-barrier energies was attributed to
nuclear incompressibility effects at the nuclear overlap and in Ref.~\cite{IH07b} to the modifications
of the inner turning point in comparison to standard folding model calculations using the frozen-density
approximation. As we shall see below, both of these assertions may be correct. In order to investigate
the dependence on incompressibility we have repeated our calculations using an older Skyrme parametrization,
SIII~\cite{BF75}, having a nuclear matter incompressibility coefficient of 356~MeV, as opposed to
most of the modern Skyrme parametrizations that have nuclear matter incompressibility around 230~MeV.
Although the SIII interaction accurately reproduces many spectroscopic properties of spherical and
close to spherical nuclei it is known not to correctly reproduce fission barriers and other phenomena
involving large deformations. In Fig.~\ref{fig:SIII} the black dashed line corresponds to the DC-TDHF calculations
using the SIII interaction. As we see, this interaction underestimates the fusion cross sections at intermediate
energies but does relatively well at low and high energies.
Although it is difficult to disentangle the
effect of incompressibility from all the other aspects of the effective interaction, one may
conclude that a new Skyrme force with a higher than the accepted nuclear matter incompressibility
of 230~MeV and fitted to reproduce large amplitude collective phenomena could better reproduce the
fusion cross sections.

In order to better examine the evolution of the nuclear density, in Fig.~\ref{fig:time-evol} we have
plotted the nuclear density at four special internuclear distances $R$ for the
$\beta_1=\beta_2=90^{\circ}$ initial orientation. Frame (a) corresponds to the nuclear
density at the outer turning point of the ion-ion potential ($R=13.1$~fm) at
$E_{\mathrm{c.m.}}\leq 86$~MeV. One striking observation from this frame is that the orientation
of the nuclear core seems to be rotated in comparison to the total nuclear density by $\pi/2$.
Frame (b) of Fig.~\ref{fig:time-evol} shows the total density at the ion-ion separation of
$R=11.0$~fm, which approximately corresponds to the location of the inner turning point at
$E_{\mathrm{c.m.}}\leq 86$~MeV. The orientation of the nuclear core is still the same as at
the outer turning point. At around $R=10$~fm the nuclear core rotates and aligns with the
total nuclear density. In frame (c) we show this at $R=9.5$~fm. The last frame (d) shows
the nuclear density at the potential minimum which occurs around $R=8.0$~fm.
Based on the above observation we can make a conjecture that while at higher energies
the potential barrier is largely determined by the nuclear surface at deep barrier
energies core nucleons play a significant role in barrier dynamics. This may manifest
itself as repulsive core in constructing a potential model for the problem.
In our calculations we can incorporate this effect by using the alignment angles of
the core rather than the total density in the calculation of the cross sections
at the lowest energies. This is shown by the black solid curve in Fig.~\ref{fig:SIII}. As
we see the results are in excellent agreement with the data at low energies.

\section{\label{sec:summary}Conclusions}
As we investigate fusion reactions involving neutron rich and deformed nuclei it is
apparent that an understanding of the structure of these nuclei is crucial to the
description of the reaction dynamics. For these nuclei various effects, such as inelastic
excitations, particle transfer, and other dynamical effects lead to substantial modification
of the naive potential barrier calculations which assume an inert core and no dynamics.
Consequently, the definition of {\it sub-barrier} fusion becomes ambiguous since
it is difficult to determine the barrier a priori.

We have performed density-constrained TDHF calculations of fusion cross sections for
the $^{64}$Ni+$^{64}$Ni system. Our results agree well with the measured data
despite having no adjustable parameters. This indicates that many
of the reaction dynamics are included in the TDHF evolution of the nuclear density.
We have also investigated the dependence of our results on the microscopic effective
interaction. We find that while all of the modern Skyrme parametrizations show
very small deviations in the fusion cross sections, which was also observed in a
more systematic study of spherical systems~\cite{UO06}, the older parametrizations
yield very different results. Since older Skyrme parametrizations were fitted mostly
to reproduce properties of spherical nuclei they may not be a good candidate for fusion
studies. On the other hand it may be desirable to investigate the dependence on incompressibility
using a more modern Skyrme parametrization, which is not yet available.

We have further investigated fusion cross sections at deep sub-barrier energies.
In the absence of a true many-body tunneling approach to nuclear fusion it is
difficult to envision the dynamical formation of the potential barrier at very deep
sub-barrier energies.
As we go further down in energy the inner turning point of the ion-ion potential
involves larger overlaps between the participating nuclei. Consequently, core nucleons
may play a more dominant role in dynamically building up the potential barrier.
In the case of $\beta_1=\beta_2=90^{\circ}$ corresponding to the lowest potential
barrier we observed that the nuclear core has a different orientation from the total
nuclear density. Based on this observation we have speculated that at the lowest
energies it may make sense to use the orientation of the core rather than the nuclear
surface. This core polarization
effect allows us to reproduce the experimental cross sections at the lowest energies.

\begin{acknowledgments}
This work has been supported by the U.S. Department of Energy under grant No.
DE-FG02-96ER40963 with Vanderbilt University.
\end{acknowledgments}


\begin{thebibliography}{99}

\bibitem{DOE02} {\it Opportunities in Nuclear Science, A Long-Range Plan for the Next
                     Decade}, DOE/NSF Nuclear Science Advisory Committee, April 2002,
                     www.sc.doe.gov/production/henp/np/nsac/nsac.html.
\bibitem{Ji04}  C. L. Jiang {\it et al.}, Phys. Rev. Lett. {\bf 93}, 012701 (2004).
\bibitem{BS82}  M. Beckerman, M. Salomaa, A. Sperduto, J. D. Molitoris, and A. DiRienzo, Phys. Rev. C {\bf 25}, 837 (1982).
\bibitem{Esb04} H. Esbensen, Prog. Theor. Phys. Suppl. {\bf 154}, 11 (2004).
\bibitem{Esb05} H. Esbensen, Phys. Rev. C {\bf 72}, 054607 (2005).
\bibitem{Esb06} \c{S}. Mi\c{s}icu and H. Esbensen, Phys. Rev. Lett. {\bf 96}, 112701 (2006).
\bibitem{IH07a} Takatoshi Ichikawa, Kouichi Hagino, and Akira Iwamoto, Phys. Rev. C {\bf 75}, 057603 (2007).
\bibitem{IH07b} Takatoshi Ichikawa, Kouichi Hagino, and Akira Iwamoto, Phys. Rev. C {\bf 75}, 064612 (2007).
\bibitem{Ba74}  R. Bass, Nucl. Phys. {\bf A231}, 45 (1974).
\bibitem{Ba80}  R. Bass, {\it Nuclear Reactions with Heavy Ions}, (Springer Verlag, New York, 1980).
\bibitem{BR77}  J. Blocki, J. Randrup, W. J. Swiatecki, and C. F. Tsang, Ann. Phys. (N.Y.) {\bf 105}, 427 (1977).
\bibitem{RV78}  J. Randrup and J. S. Vaagen, Phys. Lett. {\bf B77}, 170 (1978).
\bibitem{SG84}  M. Seiwert, W. Greiner, V. E. Oberacker, and M. J. Rhoades-Brown,Phys. Rev. C {\bf 29}, 477 (1984).
\bibitem{BH78}  J. R. Birkelund and J. R. Huizenga, Phys. Rev. C {\bf 17}, 126 (1978).
\bibitem{SL79}  G. R. Satchler and W. G. Love, Phys. Rep. {\bf 55}, 183 (1979).
\bibitem{BB77}  G. Bertsch, J. Borysowicz, H. McManus, and W. G. Love, Nucl. Phys. {\bf A284}, 399 (1977).
\bibitem{RO83a} M. J. Rhoades-Brown and V. E. Oberacker, Phys. Rev. Lett. {\bf 50}, 1435 (1983).
\bibitem{RO83b} M. J. Rhoades-Brown, V. E. Oberacker, M. Seiwert, and W. Greiner, Z. Phys. A {\bf 310}, 287 (1983).
\bibitem{TB84}  N. Takigawa and G. F. Bertsch, Phys. Rev. C {\bf 29}, 2358 (1984).
\bibitem{BT98}  A. B. Balantekin and N. Takigawa, Rev. Mod. Phys. {\bf 70}, 77 (1998).
\bibitem{LP84}  S. Landowne and S. C. Pieper, Phys. Rev. C {\bf 29}, 1352 (1984).
\bibitem{RP84}  M. J. Rhoades-Brown and M. Prakash, Phys. Rev. Lett. {\bf 53}, 333 (1984).
\bibitem{HR99}  K. Hagino, N. Rowley, and A. T. Kruppa, Comp. Phys. Comm. {\bf 123}, 143 (1999).
\bibitem{UO06b} A. S. Umar and V. E. Oberacker, Phys. Rev. C {\bf 74}, 021601(R) (2006).
\bibitem{UO06d} A. S. Umar and V. E. Oberacker, Phys. Rev. C {\bf 74}, 061601(R) (2006).
\bibitem{UO07a} A. S. Umar and V. E. Oberacker, Phys. Rev. C {\bf 76}, 014614 (2007).
\bibitem{CR85}  R. Y. Cusson, P. -G. Reinhard, M. R. Strayer, J. A. Maruhn, and W. Greiner, Z. Phys. A {\bf 320}, 475 (1985).
\bibitem{US85}  A. S. Umar, M. R. Strayer, R. Y. Cusson, P. -G. Reinhard, and D. A. Bromley, Phys. Rev. C {\bf 32}, 172 (1985).
\bibitem{Raw64} G. H. Rawitscher, Phys. Rev. 135, 605 (1964).
\bibitem{UO06c} A. S. Umar and V. E. Oberacker, Phys. Rev. C {\bf 74}, 024606 (2006).
\bibitem{ENSDF} Evaluated Nuclear Structure Data File (ENSDF),  National Nuclear Data Center, Brookhaven National Laboratory, http://www.nndc.bnl.gov/ensdf.
\bibitem{UO06}  A. S. Umar and V. E. Oberacker, Phys. Rev. C {\bf 73}, 054607 (2006).
\bibitem{CB98}  E. Chabanat, P. Bonche, P. Haensel, J. Meyer and R. Schaeffer, Nucl. Phys. {\bf A635}, 231 (1998); Nucl. Phys. {\bf A643}, 441(E) (1998).
\bibitem{DS04}  J. Dobaczewski, M. V. Stoitsov, and W. Nazarewicz, {\it Skyrme-HFB deformed nuclear mass table},
                ed. R. Bijker, R.F. Casten, and A. Frank, (AIP, New York, 2004), {\bf 726}, 51 (2004).
\bibitem{LR99}  G. A. Lalazissis, S. Raman, and P. Ring, At. Data Nucl. Data Tables {\bf 71}, 1 (1999).
\bibitem{BQ82}  J. Bartel, P. Quentin, M. Brack, C. Guet, and H. B. Hakansson, Nucl. Phys. {\bf A386}, 79 (1982).
\bibitem{BF75}  M. Beiner, H. Flocard, Nguyen Van Giai, and Ph. Quentin, Nucl. Phys. {\bf A238}, 29 (1975).
\end{thebibliography}
\end{document}